\documentclass[runningheads]{svmult}

\usepackage{makeidx}   
\usepackage{graphicx}  
\usepackage{subeqnar}  
\usepackage{multicol}  
\usepackage{physprbb}  
\makeindex             

\newcommand{\be}{\begin{equation}}
\newcommand{\ee}{\end{equation}}
\newcommand{\bea}{\begin{eqnarray}}
\newcommand{\eea}{\end{eqnarray}}
\newcommand {\cC}{{\cal C}}
\newcommand {\cn}{{\cal N}}

\begin{document}

\title*{``Go with the winners"-Simulations}

\toctitle{``Go with the winners"-Simulations}

\titlerunning{``Go with the winners"-Simulations}

\author{Peter Grassberger and Walter Nadler}
\authorrunning{Peter Grassberger and Walter Nadler}

\institute{John von Neumann - Institut f\"ur Computing,\\
     Forschungszentrum J\"ulich, D-52825 Germany}

\maketitle              

\begin{abstract}
We describe a general strategy for sampling configurations from a given
(Gibbs-Boltzmann or other) distribution. It is {\it not} based on the
Metropolis concept of establishing a Markov process whose stationary 
state is the wanted distribution. Instead, it builds weighted instances 
according to a biased distribution. If the bias is optimal, all weights 
are equal and importance sampling is perfect. If not, ``population
control" is applied by cloning/killing configurations with too high/low 
weight.  It uses the fact that nontrivial problems in statistical 
physics are high dimensional. Therefore, instances are built up in many 
steps, and the final weight can be guessed at an early stage. In 
contrast to evolutionary algorithms, the cloning/killing is done such
that the wanted distribution is strictly observed without simultaneously 
keeping a large population in computer memory. We apply this method
(which is also closely related to diffusion type quantum Monte Carlo) 
to several problems of polymer statistics, population dynamics, and 
percolation.  

\end{abstract}

\section{Introduction}

For many statistical physicists, ``Monte Carlo" is synonymous for the
Metropolis
strategy\cite{metrop} where one sets up an ergodic Markov process which has the
desired
Gibbs-Boltzmann distribution as its unique asymptotic state. There exist
numerous refinements
concerned with more efficient transitions in the Markov process (e.g. cluster
flips\cite{swendsen}
or pivot moves\cite{sokal}), or with distributions biased such that false
minima are more
easily escaped from and that autocorrelations are reduced (e.g. multicanonical
sampling\cite{berg}  and simulated tempering\cite{marinari}). But most of these
schemes remain entirely
within the framework of the Metropolis strategy.

On the other hand, stochastic simulations not based on the Metropolis strategy
have
been used from early times on. Well known examples are evolutionary (in
particular
genetic) algorithms \cite{rechenberg,holland,schwefel}, diffusion type quantum
Monte Carlo simulations
\cite{kalos,anderson,vonderlinden}, and several algorithms devised for the
simulation
of long chain molecules \cite{rosenbluth,wall,redner-rey,garel,perm}. But these
methods were
developed independently in different communities and it was not in generally
recognized
that they are realizations of a common strategy. Maybe the first who pointed
this
out clearly were Aldous and Vazirani \cite{aldous} who also coined the name
``go with
the winners". For later references who also stressed the wide range of possible
applications
of this strategy see \cite{perm-gfn,iba}. Ref. \cite{iba} points even to
applications in
lattice spin systems and Bayesian inference, fields which will not be treated
in the
present review.

As we shall see, the main drawbacks of the go-with-the-winners strategy are:
\begin{itemize}
\item The method yields correlated samples, just as the Metropolis method does.
This
makes a priori error estimates difficult \cite{aldous}. A posteriori errors,
estimated
from fluctuations of measured observables, are of course always possible. But
they can
be very misleading when sampling is so incomplete that
the really large fluctuations have not yet been seen. However, there is also a
positive
side: more easily than in the Metropolis case one can estimate whether this has
happened, and whether, therefore, the method gives reliable results or not.
\item Efficiency is not guaranteed. The go-with-the-winners strategy allows a
lot of
freedom with respect to implementation details, and its efficiency depends
on a good choice of these.
Thus, there are cases where it has not yet been successful at all,
while there are other problems where its
efficiency is not nearly matched by any other method we aware of.
On the other hand, the flexibility
of the general strategy represents a strong positive point.
\end{itemize}

Instead of giving a formal definition of the go-with-the-winners strategy, we
shall
present an example from which the basic concepts will become clear. In later
sections we shall then see how these concepts are implemented in detail and how
they are applied to other problems as well.

\section{An Example: A Lamb in front of a Pride of Lions}

The example is a very idealized problem from population dynamics (or chemical
reactions,
if you whish) \cite{bramson,redner-krip}: consider a `lamb', represented by a
random walker on
a 1-dimensional lattice $x=\ldots -1,0,1 \ldots$ with discrete time and hopping
rate $p$ per time unit, leading to a diffusion constant $D_{\rm lamb}$. It
starts
at time $t=0$ at $x=0$. Together with it, there start also $N$ `lions', $n_L$
of
them at $x_i=-1 \;(i=1,\ldots n_L)$ and $n_R=N-n_L$ at $x_i=+1$.
They also perform random walks, but with a diffusion
constant $D_{\rm lion}$ which may differ from $D_{\rm lamb}$.
Two lions can jump onto the same site without interacting with each other. But
if
a lion and the lamb meet at the same site, the lamb is eaten immediately, and
the
process is finished. Note that both the lamb and the lions are absolutely
short-sighted and stupid: there is no evading or chasing. It is for this reason
that
the model can also be interpreted as the caption of a diffusing molecule by
diffusing adsorbers.

The survival probability $P(t)$ of the lamb up to time $t$ can be estimated
easily
for a single lion, $N=1$. In this case the relative distance makes a random
walk
with diffusion constant $D_{\rm lamb}+D_{\rm lion}$ which starts at $\Delta x
=1$,
and $P(t)$ is equal to the probability that the walk has not yet hit an
absorbing
wall at $\Delta x =0$. This probability is well known to decrease as
$t^{-1/2}$, thus
\be
     P_{N=1} \sim t^{-1/2}.
\ee

The problem is less trivial but still solvable for $N=2$,
(see \cite{redner-krip} and the literature quoted there). One finds again a
power law
\be
     P_{N=2} \sim t^{-\alpha_2},
\ee
but with an exponent which depends on the ratio $D_{\rm lamb}/D_{\rm lion}$ and
on whether both lions are on the same side or on different sides of the lamb.
For the first case one gets
\be
    \alpha_2 = \left[2-{2\over \pi} \arccos{D_{\rm lamb}\over D_{\rm
lamb}+D_{\rm lion}}
            \right]^{-1} \qquad N=n_R=2, n_L=0.              \label{alpha_2}
\ee
For the case with both lions on opposite sides one obtains a similar
expression.
If $D_{\rm lamb}=D_{\rm lion}$, Eq.~(\ref{alpha_2}) reduces to $\alpha_2 =
3/4$.

For any $N>2$ one can still prove rigorously that asymptotically holds, for large
$t$,
\be
     P_N \sim t^{-\alpha_N},    \label{alpha_N}
\ee
but this time one cannot give closed expressions for $\alpha_N$. Numerical
values
for the case where all lions are on one side have been obtained for several $N$
by
direct simulation ($\alpha_3 \approx 0.91, \;\alpha_4 \approx 1.03,
\;\alpha_{10}
\approx 1.4$ \cite{bramson}), but these estimates become more and more
difficult for increasing
$N$ because of the exceedingly small chance for the lamb to survive
sufficiently
long to allow precise measurements.

Such numerical estimates would be welcome in order to test an asymptotic
estimate
for $N\to\infty$ \cite{redner-krip}. In this limit, the location of the
{\it outmost} of a group of lions moves nearly deterministically: If the lion
who made the front at time $t$ lags behind, there will be another lion who
overtakes
him, so that the front continues to move on with maximal speed. Assuming that
the fluctuations in the motion of the front can be neglected, the authors of
\cite{redner-krip} found
\be
     \alpha_N \approx {D_{\rm lion}\over 4D_{\rm lamb}} \ln N \qquad N=n_R\gg
1, n_L=0.
                                                                 \label{alpha_infty}
\ee
In the same spirit, the optimal strategy of a lamb squeezed between $N/2$ lions
to its left and $N/2$ to its right would be to stand still. Assuming that this
single path dominates in the limit $N\to\infty$, one finds simply $\alpha_N
\approx N$.

As we said, straightforward simulations to check these predictions are
inefficient.  In order to improve the efficiency, one can think of two tricks:

\fbox{ {\bf Trick 1 : } Make occasional ``enrichment" steps. }

\noindent
In particular, this might mean that one
starts with $M \gg 1$ instances. As soon as the number of surviving instances
has decayed to a number $<M/2$, one makes a clone of each instance (note that 
lambs can be cloned also in reality, but on the computer we clone the entire
configuration
consisting of lamb and lions!). This boosts the number of instances again up to
$\approx M$, and one can repeat the game. One has just to remember how often
the sample had been enriched when computing survival probabilities, i.e.
each instance generated carries a relative statistical weight $w=1/2^c$,
with $c$ the number of cloning steps.

\fbox{ {\bf Trick 2 : } Replace the random walks by {\it biased} random walks}

\noindent
Not only should the lamb preferentially run away from the lions, but also 
the lions should run away from the lamb in 
order to obtain long-lived samples that contribute to Eq.~(\ref{alpha_N}).
If just this were done without compensation, this would of course give wrong
results. But we can correct for this bias by giving {\it weights} to each
instance.  For each step (of either the lamb or the lion) that was made 
according to a biased pair of probabilities $\{p_L,p_R\}$
instead of $\{p,p\}$, we should multiply the weight by a factor $p/p_R$ if
the actual step was to the right, and by a factor $p/p_L$ if the step was to
the left. In this way the weights compensate exactly, on average, the fact that
not all walks were sampled with the same probability. This trick is indeed very
general.
In any sampling procedure where some random move should be done with
probability $p>0$
in order to obtain an unbiased sample, one can replace $p$ by any other
probability
$p'\ne0$ if we at the same time use weighted samples and multiply the current
sample weight by $p/p'$.

Actually, in view of the second trick, the first one is clearly not optimal.
Instead of cloning irrespective of its weight, one would like to clone
preferentially
those configurations which have high weight. Thus we replace the first trick by

\fbox{ {\bf Trick 1' : } Clone only configurations with high weight.}

\noindent
Choose a cloning threshold $W_+(t)$. 
It can be in principle an arbitrary function
of $t$, and it need not be kept fixed during the simulation; thus it can be
optimized on-line. Good choices will be discussed later. If a configuration
at some time $t$ has weight $w>W_+(t)$, it is cloned and both clones are given
weight $w/2$.

On the other hand, a too strong bias and too frequent cloning could result in
configurations which have too small weight. Such configurations are just costly
in terms of CPU time, without adding much to the precision of the result. But
we are not allowed to kill (``prune") them straigh away, since they do carry
some weight nevertheless. Instead, we use

\fbox{ {\bf Trick 3 : } Kill probabilistically configurations with low weight. }

\noindent
Choose a pruning threshold $W_-(t)$.
The same remarks apply to it as to
$W_+(t)$. If $w<W_-(t)$, we call a random number $r$ uniformly in $[0,1]$.
If $r<1/2$, we prune. Otherwise, if $r>1/2$, we keep the configuration and
double
its weight, $w\to 2w$. Again this does not introduce a bias, as far as {\it
averages}
are concerned.

In principle, that's all. One can modify the tricks {\bf 1'}, {\bf 2} and {\bf 3} by making more
than one clone at each enrichment step, by killing with probability $\neq 1/2$,
or by letting $W_\pm$ depend also on other variables. Whether such further
improvements are helpful will depend on the problem at hand, in the case
of lamb \& lions it seems they were not. Indeed, in this problem also pruning
was not needed if the bias was not too strong, but this is somewhat special.

\begin{figure}
\begin{center}
\includegraphics[width=.34\textwidth,angle=270]{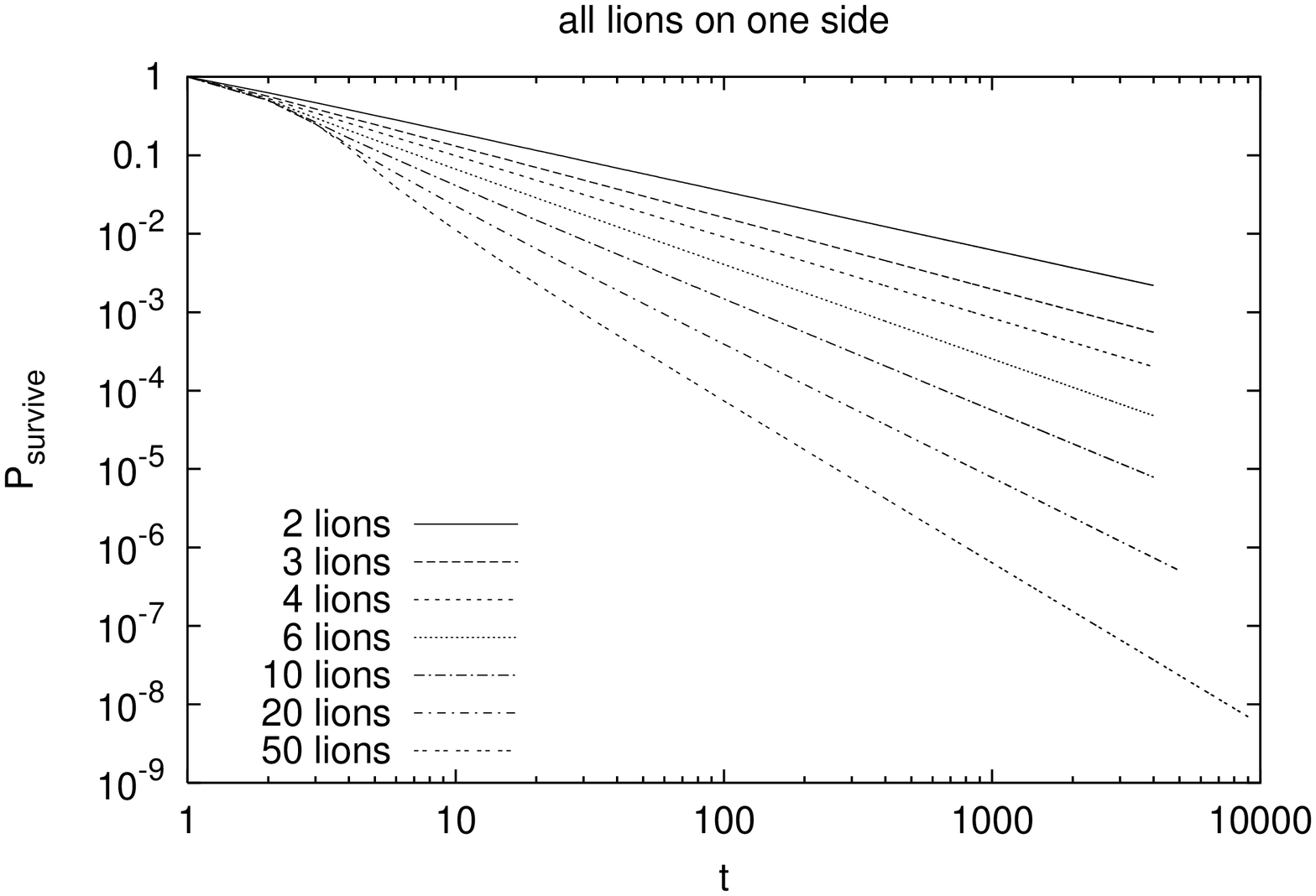}
\includegraphics[width=.34\textwidth,angle=270]{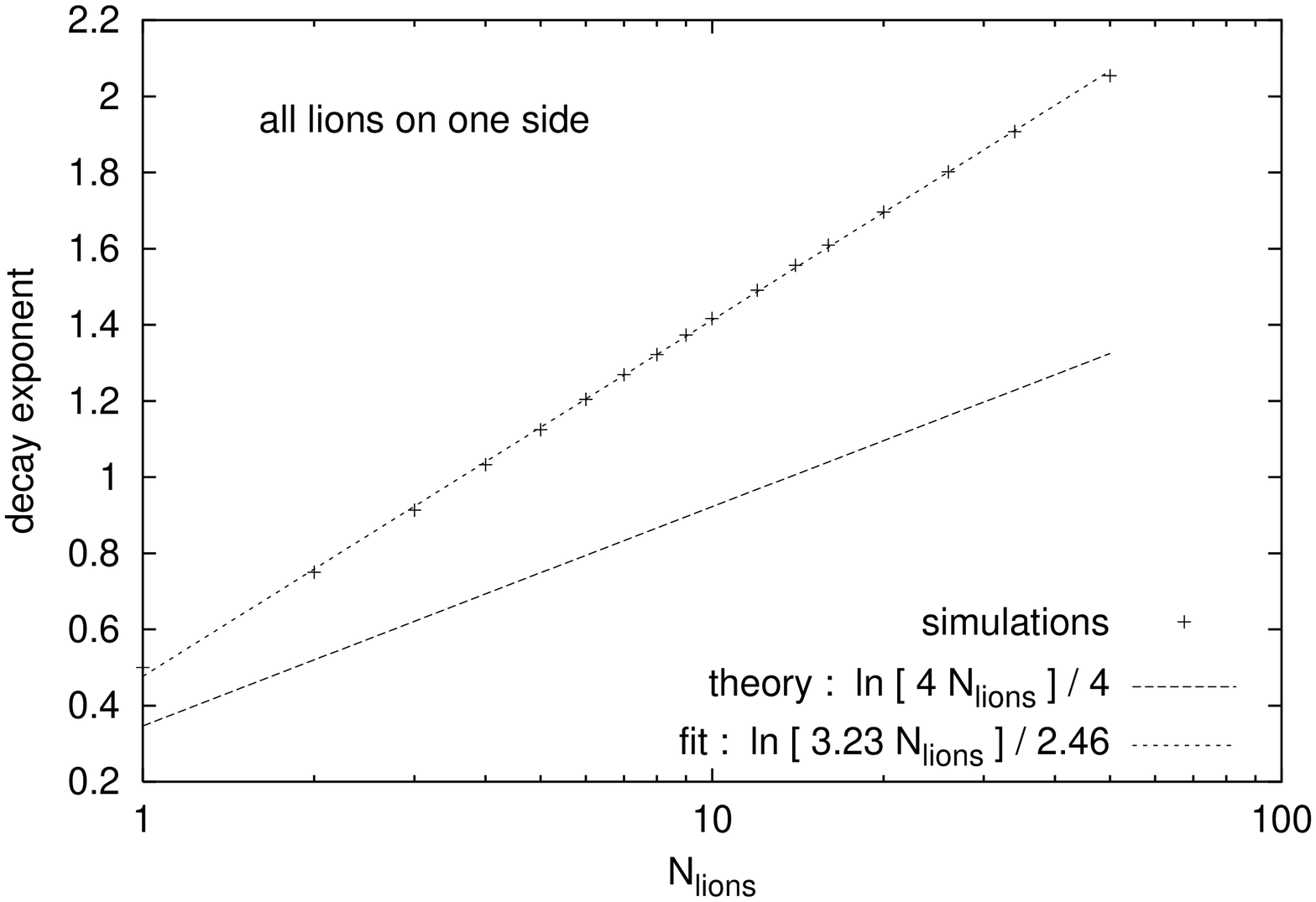}
\end{center}
\caption[]{Left panel: Survival probabilities for a lamb starting next to $N$
lions,
    all of whom are on the same side.
    Lamb and lions both make ordinary random walks with
    $D_{\rm lion}=D_{\rm lamb}=1/2$. Right panel: Corresponding decay
exponents.
    The lower dashed line represents the prediction from
Eq.~(\ref{alpha_infty}).}
\label{figs12}
\end{figure}

\begin{figure}
\begin{center}
\includegraphics[width=.34\textwidth,angle=270]{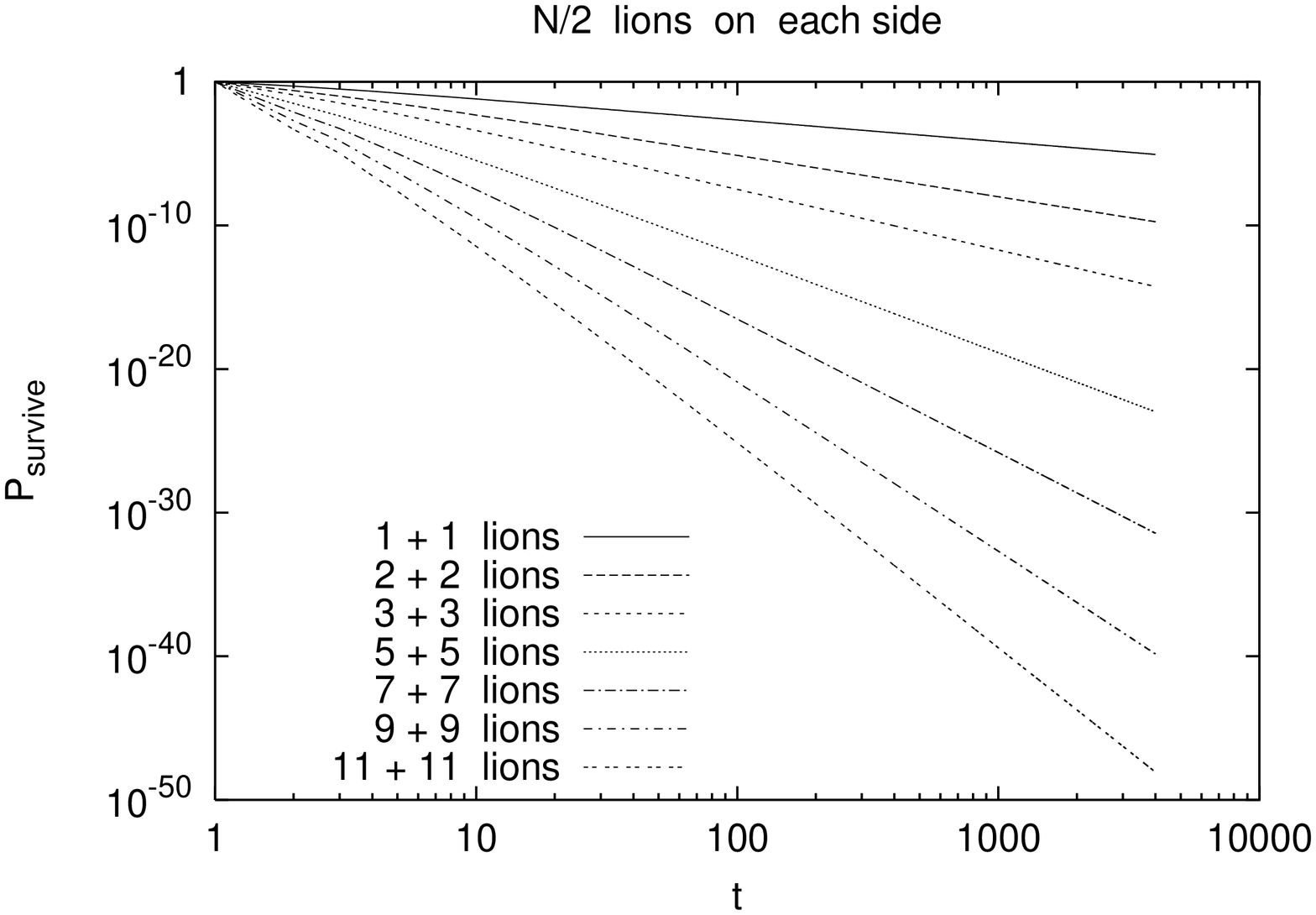}
\includegraphics[width=.34\textwidth,angle=270]{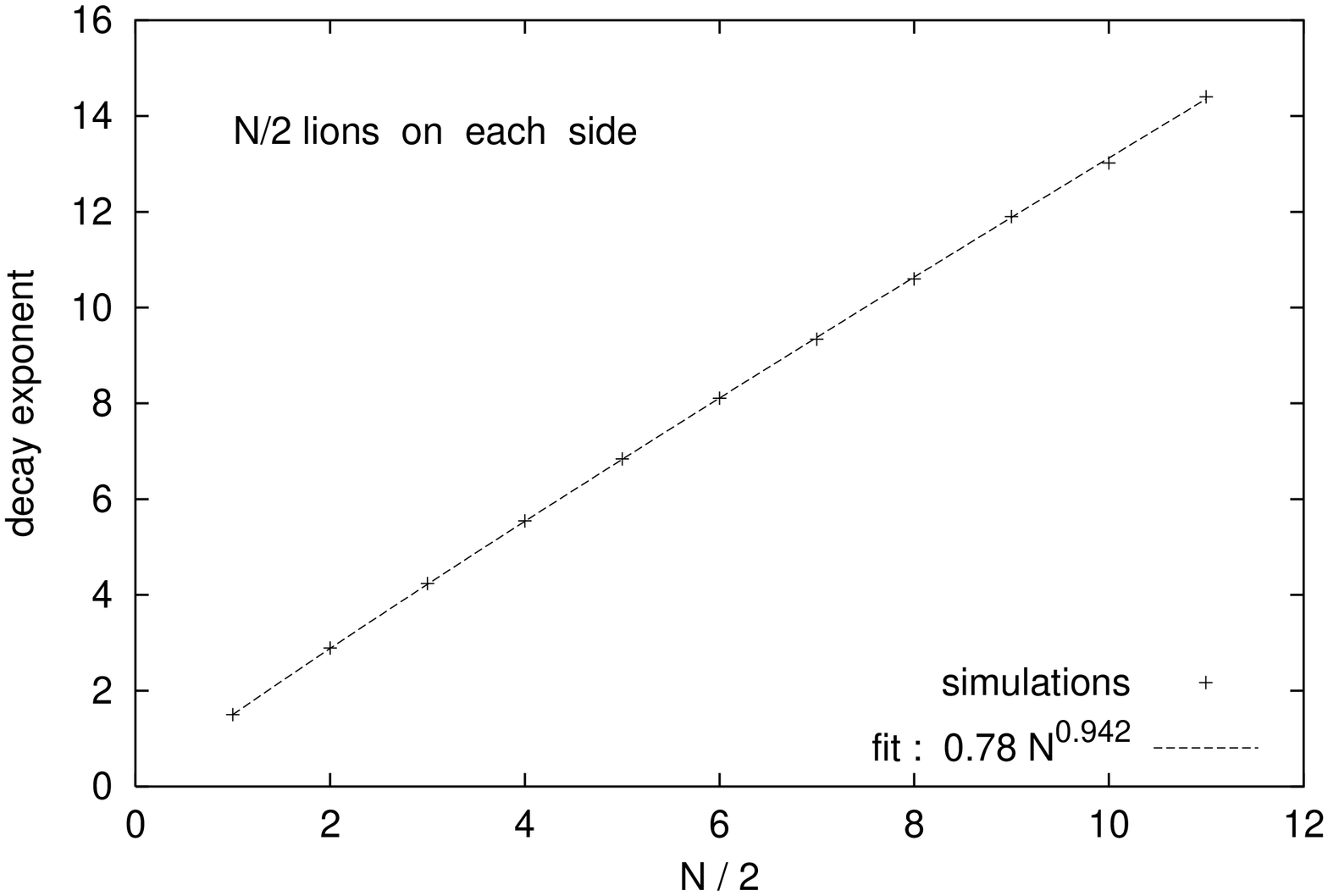}
\end{center}
\caption[]{Same as Fig.1 but for $N/2$ lions on each side of the lamb.
    This time neglecting fluctuations of the
    front of the group of lions would give $\alpha_N = N$.}
\label{figs34}
\end{figure}

Before going on and describing the detailed implementation, let us just see
some results. Probabilities $P_N(t)$ for all lions at the same side and the
resulting decay exponents are shown in Fig.1, for $N$ up to 50. We see
that Eq.~(\ref{alpha_infty}) is qualitatively correct in predicting a
logarithmic
increase of $\alpha_N$, but not quantitatively. Obviously, fluctuations of the
front of the pride of lions are not negligible. The data on the right panel
show
a slight downward curvature. This might be an indication that
Eq.~(\ref{alpha_infty})
is asymptotically correct, but then asymptotia would set in only at extremely
large
values of $N$. The same conclusion is
reached when $N/2$ lions are on either side, as seen from Fig.2. In that case
the raw data, shown in the left panel, clearly demonstrate the power of the
algorithm:
We are able to obtain reliable estimates of probabilities as small as
$10^{-50}$,
which would have been impossible with straightforward simulation.

\section{Other Examples}

\subsection{Multiple Spanning Percolation Clusters}

\begin{figure}[b]
\begin{center}
\includegraphics[width=.45\textwidth,angle=270]{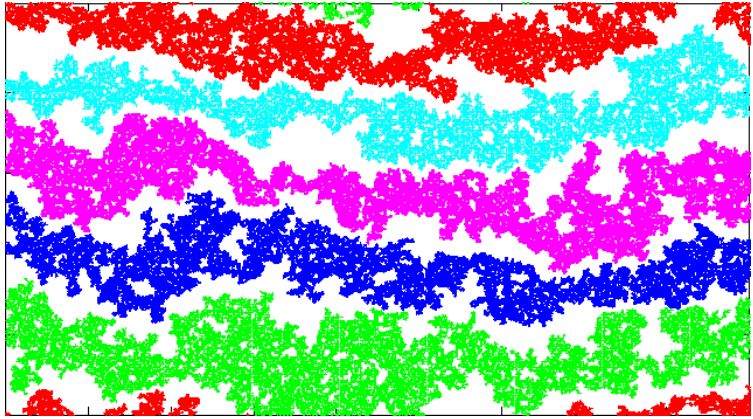}
\end{center}
\caption[]{Configuration of 5 spanning site percolation clusters on a
    lattice of size $500\times 900$. Any two clusters keep a distance of at
    least 2 lattice units. Lateral boundary conditions are periodic.
    The probability to find 5 such spanning clusters
    in a random disorder configuration is $\approx 10^{-92}$.}
\label{figs5}
\end{figure}

Let us now consider percolation \cite{stauffer} on a large but finite
rectangular
lattice in any dimension $2\leq d < 6$. We single out one direction as
``spanning
direction". In this direction boundary conditions are open (surface sites just
have no neighbours outside the lattice), while boundary conditions in the other
direction(s) might be either open or periodic.
Up to some six years ago there was a general believe, based on a misunderstood
theorem, that there is at most one spanning cluster in the limit of large
lattice
size, keeping the shape of the rectangle fixed ($L_i = x_i L,\; L\to\infty,\;
i=1,
\ldots d$). A `spanning cluster' is a cluster which touches both boundaries
in the spanning direction.

Since there is no spanning cluster for subcritical percolation (with
probability decreasing exponentially in the lattice size $L$), and since there
is exactly one in the supercritical case, the relevant case is only critical
percolation. For that case it is now known that the probabilities $P_k$ to have
exactly $k$ spanning clusters are all non-zero in the limit $L\to\infty$. In
two dimensions they were calculated by Cardy exactly using conformal invariance
\cite{cardy}, but
in dimensions $\geq 3$ no exact results are known. The only analytical `result'
is a conjecture by Aizenman \cite{aizenman}, stating that for a lattice
of size $L\times \ldots \times L \times (rL)$ ($rL$ is the length in the
spanning
direction)
\be
   P_k \sim e^{-\alpha r}        \label{k-cluster}
\ee
with
\be
   \alpha \propto k^{d/(d-1)} \qquad {\rm for} \;\; k\gg 1.   \label{aizen}
\ee

\begin{figure}[b]
\begin{center}
\includegraphics[width=.34\textwidth,angle=270]{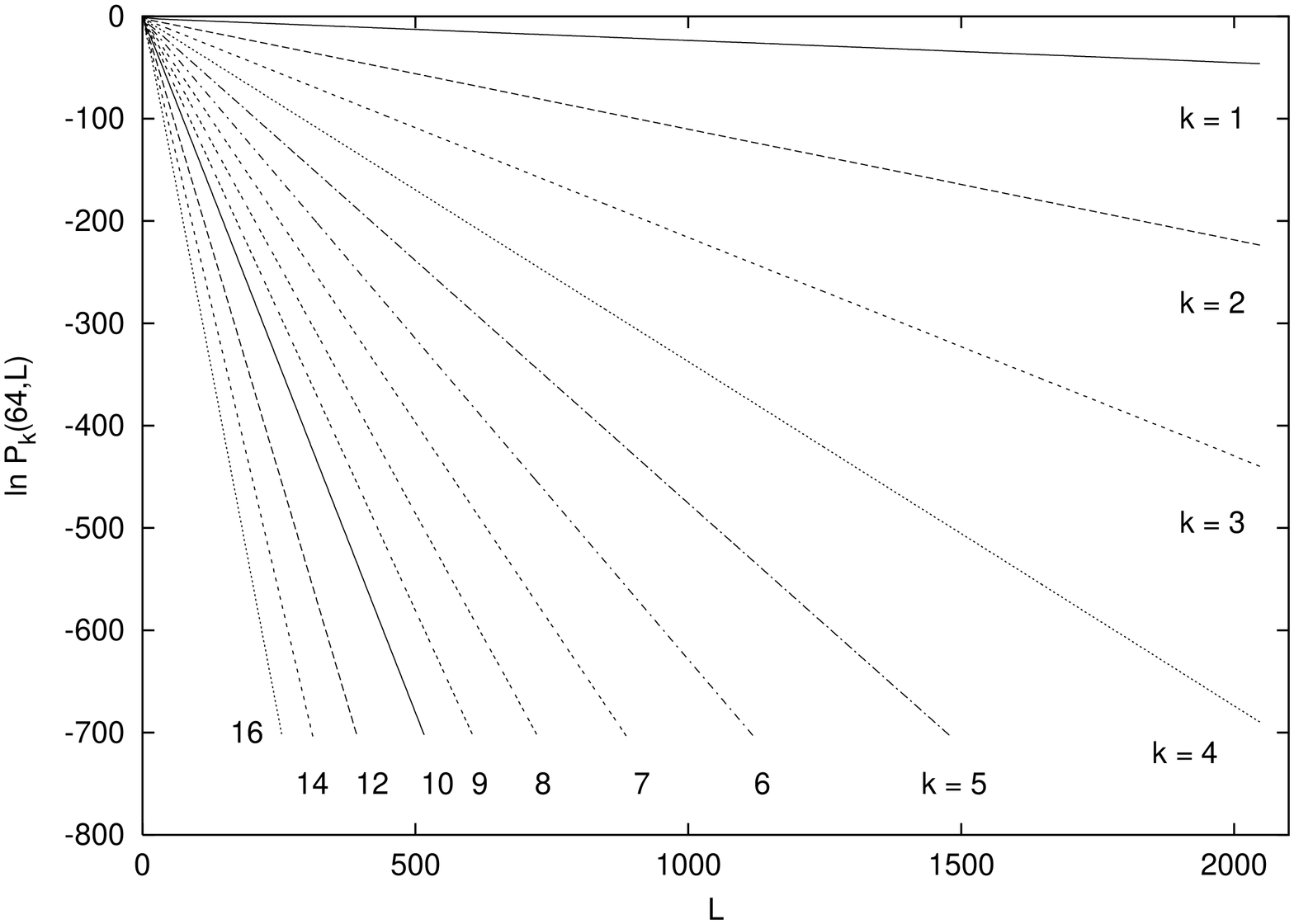}
\includegraphics[width=.34\textwidth,angle=270]{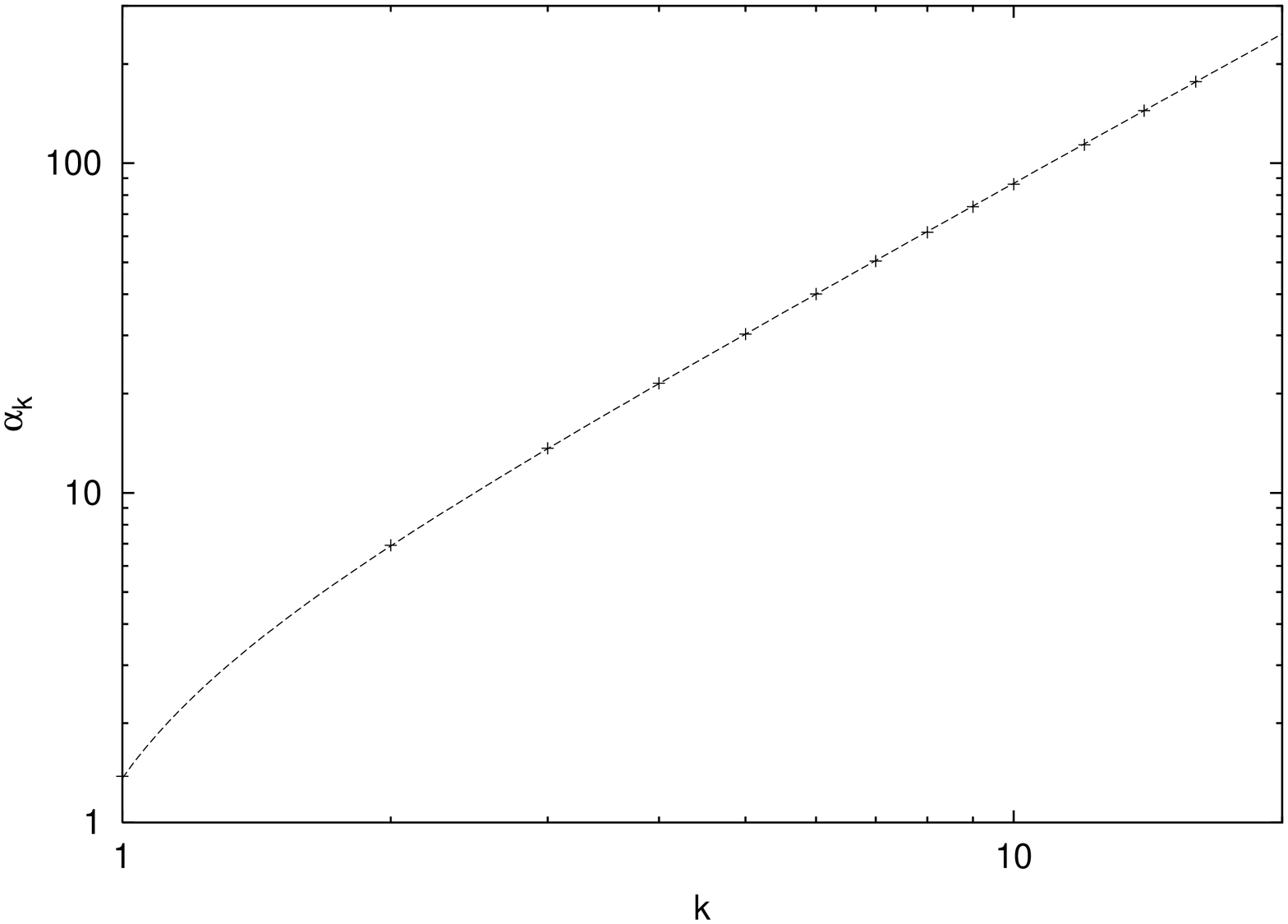}
\end{center}
\caption[]{Left panel: probabilities to have $k$ spanning clusters on simple
    cubic lattices of size $64\times 64\times L$, with $k \le 16$ and $L\le
2000$.
    Right panel: Decay exponents $\alpha_k$ versus $k$ obtained from the data
    on the left panel, and from similar data with transverse lattice sizes
    $16\times 16$, $32\times 32$, and $128\times 128$. The dashed line is a
    fit with $\alpha_k \sim k^{3/2}$ for large $k$.}
\label{figs67}
\end{figure}

Cardy's formula in $d=2$ agrees with Eqs.~(\ref{k-cluster}) and (\ref{aizen}),
for periodic
transverse boundary conditions it is
\be
   \alpha = {2\pi\over 3}\left(k^2-{1\over 4}\right) \qquad k\geq 2,\; d=2\;.
                                     \label{cardy}
\ee
It has recently been generalized \cite{aizen-dupl-ahar} to the case where the
clusters are seperated by at least two lattice units (i.e., there are at least
two non-intersecting paths on the dual lattice between any two clusters). In
that case
\be
   \alpha = {2\pi\over 3}\left(({3k\over 2})^2-{1\over 4}\right) \;.
                                     \label{ada}
\ee
In order to test Eqs.~(\ref{aizen})-(\ref{ada}) for a wide range of values of
$k$ and $r$, one has to simulate events with tiny probabilities,
$\ln P_k \sim -10^2$ to $ -10^3$. It is thus not surprising that previous
numerical studies have verified Eq.~(\ref{cardy}) only for small values of
$k$, and have been unable to verify or disprove Eq.~(\ref{aizen})
\cite{sen,shchur}.

In order to demonstrate that such rare events can be simulated efficiently
with the go-with-the-winners strategy, we show in Fig.3 a rectangular lattice
of size $500\times 900$ with 5 spanning clusters which keep distances $\geq 2$.
Eq.(\ref{ada}) predicts for it $P_k = \exp(-336\pi/5) \approx 10^{-92}$.
This configuration was obtained by letting 5 clusters grow simultaneously,
using a standard cluster growth algorithm \cite{leath}, from the left border.
Precautions were taken that they grow with the same speed towards the right,
i.e. if one of them lagged behind, the growth of the others was stopped until
the lagging cluster had caught up. If one of them died, or if two came closer
than two lattice units, the entire configuration was discarded. If not, cloning
was done as described in trick {\bf 1'}. Note that here the growth was made without
bias (it is not obvious what this bias should have been), and therefore
the weight was just determined by the cloning. Due to that and since there are 
no Boltzmann weights, no configuration could get too high a weight, and therefore
no pruning was necessary either.

In this way we could check Eqs.~(\ref{cardy}),(\ref{ada}) with high precision.
We do not show these data. Essentially they just test the correctness of
our algorithm.

More interesting is the test of Eq.~(\ref{aizen}) for $d=3$. Estimated
probabilities for up to 16 spanning clusters, on lattices of sizes up to
$64\times 64\times 2000$, are shown on the left panel of Fig.4. Note that
probabilities now are as small as $10^{-300}$. Values of $\alpha$ obtained
from these simulations and from similar ones at different lattice sizes (in
order to eliminate finite-size corrections) are shown in the right panel
of Fig.4. The dashed line there is a fit \cite{gra-zif}
\be
   \alpha = 2.76(k^2-0.61)^{3/4}
\ee
Even if we should not take this fit too serious, we see clearly that $\alpha
\propto k^{3/2}$ for $k\to\infty$, in perfect agreement with Eq.(\ref{aizen}).

\subsection{Polymers}

Another big class of problems where the go-with-the-winners strategy is
naturally
applied are configurational statistics of long polymer chains. It is well known
that linear polymers in good solvents form random coils which differ from
random walks by having size
\be
   R \sim N^\nu
\ee
with $\nu \neq 1/2$: $\nu=3/4$ in 2 dimensions, and $\nu\approx 0.588$ in $d=3$
\cite{degennes,cloizeaux}. The canonical model which gives this anomalous
scaling
is the self avoiding random walk. Anomalous scaling laws in other universality
classes are obtained by attaching polymers to impenetrable boundaries, to
attractive walls, by adding monomer-monomer attraction, etc. Simulating long
chain
molecules was thus a vigorous problem since the very early days of electronic
computers.

The most straightforward method to simulate a self avoiding random walk on a
regular lattice is to start from one end and to make steps in random
directions.
As long as no site is visited twice, every configuration should have the same
weight. But as soon as a site is visited which has already been visited before,
the energy becomes infinite because of hard core repulsion, the weight thus
becomes zero, and the configuration can be discarded. This leads to exponential
``attrition"  -- the number of generated configurations of length $t$ decreases
as $C(t)\sim \exp(-at)$ -- and to a very inefficient code.

A first proposal to avoid -- or at least reduce -- this attrition was made by
Rosenbluth and Rosenbluth \cite{rosenbluth}. They proposed to bias the
sampling by replacing steps to previously visited sites by steps to unvisited
ones, if possible. Take e.g. a simple cubic lattice. Except for the very first
step, there will be at most 5 free neighbours for the next move. If there is
no free neighbour at any given moment, the configuration must be discarded.
Otherwise, if there are $m\geq 1$ free neighbours, one selects one of them
randomly and moves to it. At the same time, in order to compensate for this
bias one multiplies the weight of the configuration by a factor $\propto m$
(the value of the proportionality constant is irrelevant for estimates of
averages,
and affects the partition sum in a trivial way only).

Although this allows much longer chains to be simulated, the Rosenbluth method
is far from perfect because it leads to very large weight fluctuations
\cite{batoulis}. As an alternative, enrichment was therefore proposed -- in
the form of trick {\bf 1} of Sec.2 -- in \cite{wall}. But more efficient than either
is the full go-with-the-winners strategy with all three steps {\bf 1'}, {\bf 2}, and {\bf 3}.
Population control (pruning/cloning) is of course done on the basis of full
statistical weights, including both Boltzmann and bias correction factors.
This was first used in \cite{garel} and later, with a different implementation,
in \cite{perm}. In the latter, it was called the `pruned-enriched Rosenbluth
method' (PERM).

PERM is particularly efficient near the so-called `theta-' or coil-globule
transition. This transition occurs when we start with a good solvent and
make it worse, e.g. by lowering the temperature $T$. The repulsive interaction
between monomers and solvent molecules lead to an effective monomer-monomer
attraction which would like to make the polymer collapse into a dense globule.
If $T$ is sufficiently high, this is outweighed by the loss of entropy
associated
to the collapse. But at $T<T_\theta$, the entropy is no longer sufficient to
prevent the collapse. According to the generally accepted scenario, the
theta-point is a tricritical point with upper critical dimension $d=3$
\cite{degennes,cloizeaux}.

At the 3-d theta-point bias correction and Boltzmann factors nearly cancel.
Therefore, long polymers have essentially random walk configurations with
very small (logarithmic) corrections. Therefore, an unbiased random walk
(with just a non-reversal bias: no 180 degree reversals are allowed) is
already sufficient to give good statistics with very few pruning and
enrichment events. In \cite{perm} chains made of up to 1,000,000 steps
could be sampled with high statistics within modest CPU time.

Applications of PERM to other polymer problems are treated in
\cite{stiff,mix,fold,parallel,cylinder,manhattan,localize,dna,hetero,annealed,hager,prellberg}.
We want to discuss here just two application, namely the `melting'
(denaturation) of DNA \cite{dna} and the low-energy ("native") states
of heteropolymers\cite{fold}.

\begin{figure}[b]
\begin{center}
\includegraphics[width=.34\textwidth,angle=270]{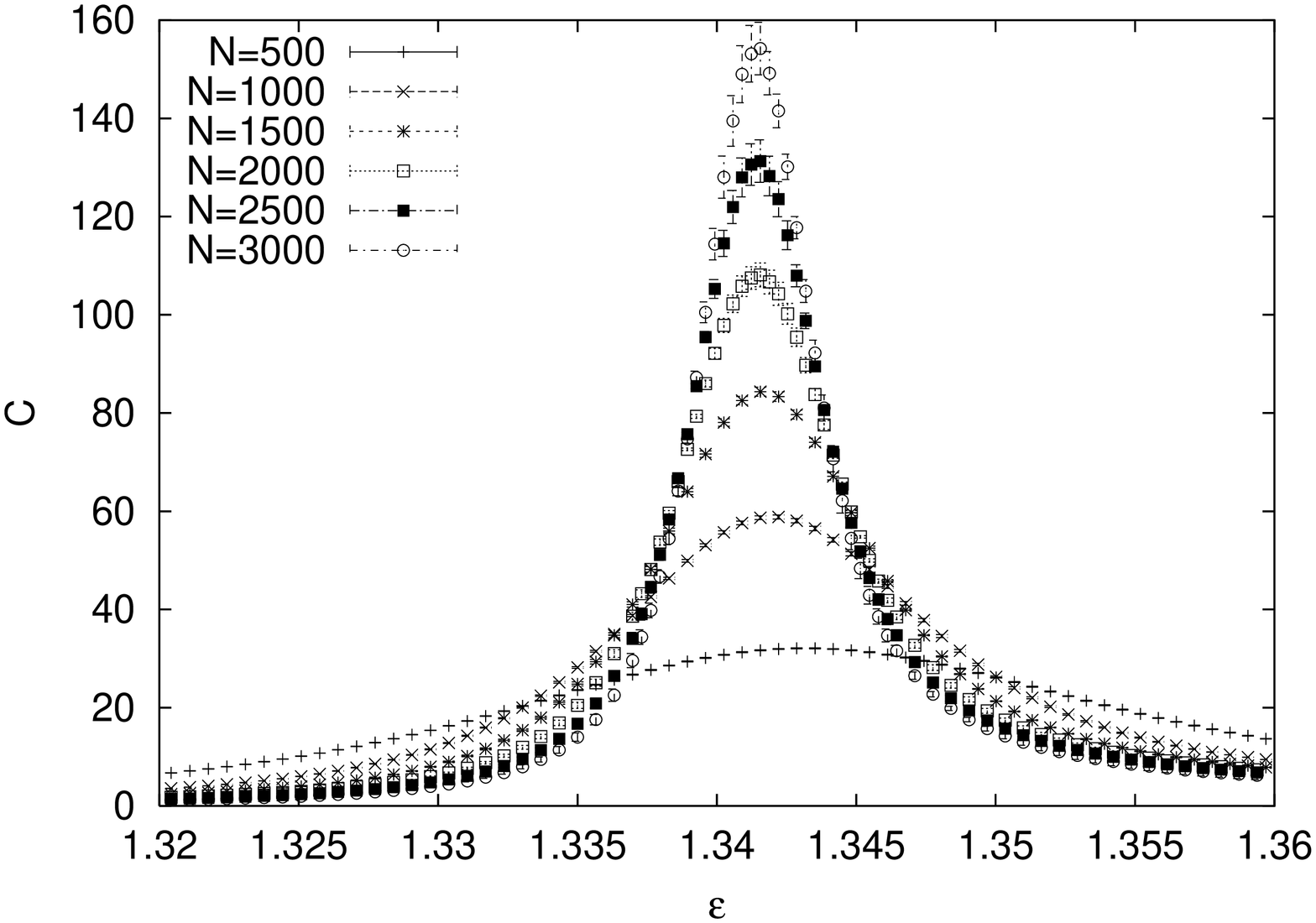}
\includegraphics[width=.34\textwidth,angle=270]{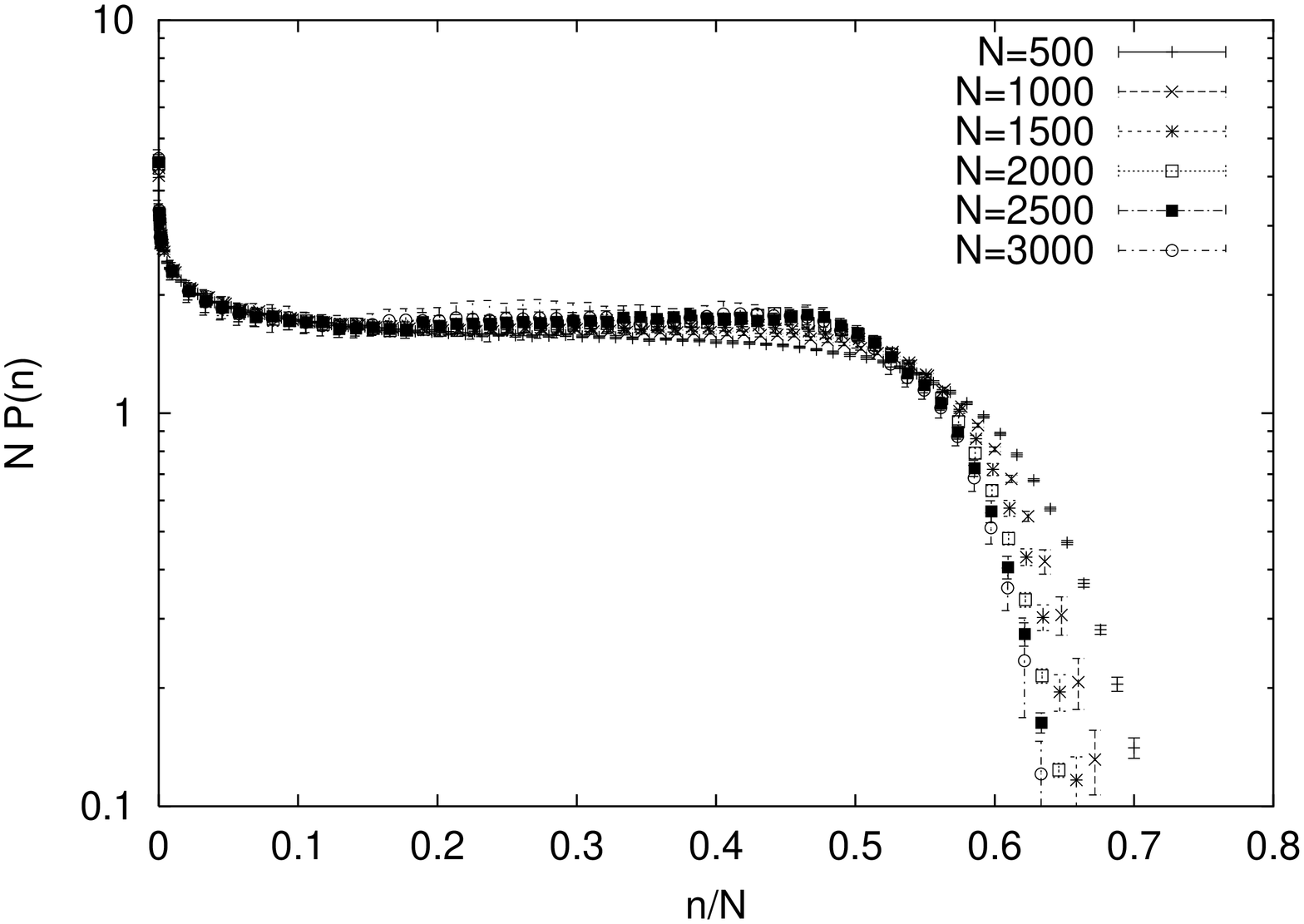}
\end{center}
\caption[]{Left panel: Specific heat as a function of $\epsilon$ at $T=1$,
    for single strand length $N=500, \ldots 3000$.
    Right panel: Histograms of the number of contacts, for the same chain
    lengths, at $\epsilon=\epsilon_c$.
    On the horizontal axis is plotted $n/N$ as is appropriate for
    a first order transition.}
\label{figs910}
\end{figure}

\begin{figure}
\begin{center}
\parbox{.50\textwidth}{\includegraphics[width=.4\textwidth,angle=270]{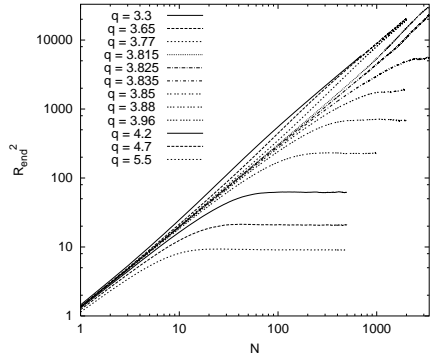}}
\parbox{.41\textwidth}{\caption[]{\sloppy Average squared end-to-end distance
$R_{\rm end}^2$ for various values of
  $q=e^\epsilon$, plotted against $N$ on a double logarithmic scale. Since all
curves
  are based on independent runs, their typical fluctuations relative to each
other
  indicate the order of magnitude of their statistical errors.}}
\end{center}
\label{fig11}
\end{figure}

\subsubsection{DNA Melting}

As is well known,
DNA in physiological conditions forms a double helix. Changing the pH value
or increasing $T$ can break the hydrogen bonds between the A-T and C-G
pairs, and a phase transition to an open coil, with higher energy but also
with higher entropy, occurs. This transition has been studied experimentally
since about 40 years. It seems to be very sharp, experimental data are
consistent with a first order transition \cite{wartell}. While a second
order transition would be easy to explain \cite{poland,fisher},
constructing models which give first order transitions turned out to be
much more difficult \cite{kafri-muk-peliti}.

The model studied in \cite{dna} lives on a simple cubic lattice. A double
strand of DNA with length $N$ is described by a diblock copolymer of
length $2N$, made of $N$ monomers of type $A$ and $N$ monomers of type $B$.
All monomers have excluded volume interactions, i.e. two monomers cannot
occupy the same lattice site, with one exception: The $k$-th $A$-monomer
and the $k$-th $B$-monomer, with $k = 1 \ldots N$ being their index counted
from the center where both strands are joint together, can occupy the same
site. If they do so, then they even gain an energy $-\epsilon$. This models the
binding of complementary bases.

The surprising result of simulations of chains with $N$ up to 4000 is that
the transition is first order, but shows finite scaling behaviour as expected
for a second order transition with cross-over exponent $\phi=1$. To demonstrate
this, we first show in Fig.5 the specific heat as a function of $\epsilon$ at
$T=1$,
for several chain lengths. We see a linear increase of the peak height with $N$
which indicates a first order transition. In the right hand panel of Fig.5 are
plotted energy histograms for the same chain lengths. Energy is measured in
terms of number $n$ of contacts, divided by chain length $N$. One sees two
maxima, one at $n=0$ and the other at $n\approx N/2$, whose distance scales
proportionally to $N$. This again points to a first order transition. But in
contrast to usual first order transitions the minimum between these two maxima
does not become deeper with increasing $N$. This is due to the absence of any
analogon to a surface tension. Finally, in Fig.6 we show average squared
end-to-end distances. They obviously diverge for infinite $N$ when the
transition point is approached from low temperatures. This is typical of a
second order transition. A more detailed analysis shows that this divergence
is $R_{\rm end} \sim (\epsilon-\epsilon_c)^\nu$, as one would expect for
a transition with $\phi=1$ \cite{dna}.

\subsubsection{Native Configurations of Toy Proteins Models}

Predicting the native states of proteins is one of the most challenging
problems in mathematical biology \cite{creighton}.
It is not only important for basic science,
but could also have enormous technological applications. At present,
such predictions are mostly done by analog methods, i.e. by comparing
with similar amino acid sequences whose native states are already known.
More direct approaches are hampered by two difficulties:
\begin{itemize}
\item Molecular force fields are not yet precise enough. Energies between
native and misfolded states are usually just a few eV, which is about the
typical precision of empirical potentials. Quantum mechanical ab initio
calculations of large biomolecules are impossible today.
\item Even if perfect force fields were available, present day algorithms
for finding ground states are too slow. One should add that the accepted
dogma is that native states -- at least of not too large proteins --
are essentially energetic ground states.
\end{itemize}

\begin{figure}[b]
\begin{center}
\includegraphics[width=.43\textwidth]{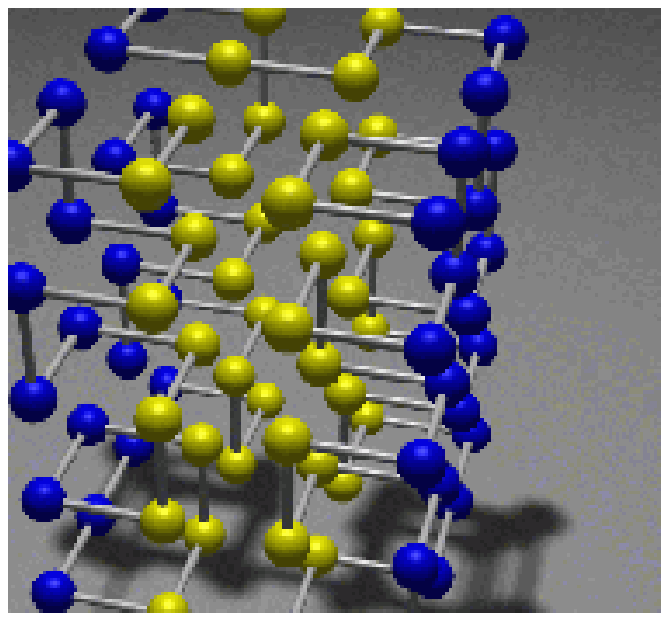}
\includegraphics[width=.43\textwidth]{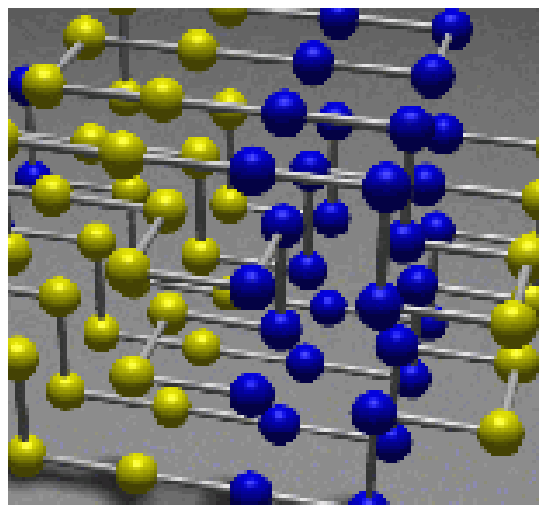}
\end{center}
\caption[]{Left: Putative native state of the ``four helix bundle" sequence
   of \cite{otoole}. It has $E=-94$, fits into a rectangular box, and consists
   of three homogeneous layers. Structurally, it can be interpreted as four
helix
   bundles. Right: True ground state with $E=-98$. Its shape is highly
symmetric
   although it does not fit into a rectangular box. It is degenerate with
   other configurations not discussed here.}
\label{fig1213}
\end{figure}

In view of the second problem, there exists a vast literature on finding
ground states of artificially constructed heteropolymers. Most of these
models are formulated on a (square or simple cubic) lattice and use only
few monomer types. The best known example is the HP model if K. Dill
\cite{dill85} which has two types of amino acids: hyrophobic (H) and
hyrophilic (polar, P) ones. With most algorithms, one can find ground states
typically for random chains of length up to $\sim 50$.

In \cite{fold} we have used PERM to study several sequences, of the HP model
and of similar models, which had been discussed previously by other authors.
In all cases we found the lowest energy states found also by these authors,
but in several cases we found new lowest energy states. A particularly
impressive example is a chain of length 80 with two types of monomers with
somewhat artificial interactions: Two monomers on neighbouring lattice sites
contribute an energy $-1$ if they are of the same type, but do not contribute
any energy if they are different \cite{otoole}. A particular sequence was
constructed such that it should fold into a bundle of four ``helices" with
an energy $-94$ \cite{otoole}. Even with a specially designed algorithm,
the authors of \cite{otoole} were not able to recover this state. With PERM
we not only found it easily, we also found several lower states, the lowest
one having energy $-98$ and having a completely different structure. Instead
of being dominanted by $\alpha$-helices, it has mostly $\beta$-sheets (as far
as these structures can be identified on a lattice), see Fig.7. Since PERM
gives not only the ground state but the full partition sum, we could also
follow the transition between mostly helical states at finite $T$ and
the sheetlike ground state. We found a peak in the specific heat associated
to this transition which could have been mistaken as a sign of a transition
between a molten globule and the frozen native state.

\subsection{Lattice Animals (Randomly Branched Polymers)}

Consider the set of all connected clusters $\cC_n$ of $n$ sites on a regular
lattice,
with the origin being one of these sites, and with a weight defined on
each cluster. The ($n$-site) lattice animal problem is defined by giving the
same
weight to each cluster. The last requirement distinguishes animal statistics
from statistics of percolation clusters. Take site percolation for
definiteness,
with `wetting' probability $p$. Then a cluster of $n$ sites with $b$ boundary
sites
carries a weight $p^n(1-p)^b$ in the percolation ensemble, while its weight in
the
animals ensemble is independent of $b$. In the limit $p\to 0$ this difference
disappears obviously, and the two statistics coincide. Due to universality, we
expect indeed that the scaling behavior is the same for any value of $p$ less
than
the critical percolation threshold $p_c$. It is generally believed that lattice
animals are a good model for randomly branched polymers \cite{lubensky}.

While there exists no simple and efficient algorithm for simulating large
animals
which also gives estimates for the partition sum, there exist very simple
and efficient algorithms for percolation clusters. The best known is presumably
the
Leath algorithm\cite{leath} which constructs the cluster in a ``breadth first"
(see next section) way.

Our PERM strategy consists now in starting off to generate subcritical
percolation
clusters by the Leath method, and in making clones of those growing clusters
which
contribute more than average to the animal ensemble \cite{animals}. Since we
work
at $p<p_c$, each cluster growth would stop sooner or later if there were no
enrichment. Therefore we do not need
explicit pruning. The threshold $W_+$ for cloning is chosen such that it
depends
both on the present animal weight and on the anticipated weight at the end of
growth.

\begin{figure}
\begin{center}
\parbox{.49\textwidth}{\includegraphics[width=.45\textwidth,angle=270]{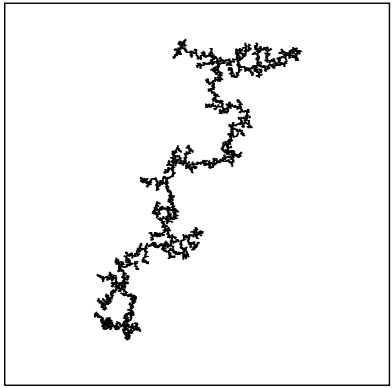}}
\parbox[t]{.42\textwidth}{\caption[]{A typical lattice animal with 8000 sites
on the square lattice.}}
\end{center}
\label{fig14}
\end{figure}

Usually, with growth algorithms like the Leath method, cluster statistics
is updated only after clusters have stopped growing. But, as outlined below,
one
can also include contributions of still growing clusters. For percolation, this
reduces slightly the statistical fluctuations of the cluster size
distributions,
but the improvement is small. On the other hand, this improved strategy is
crucial
when using PERM to estimate animal statistics.

Consider a growing cluster during Leath growth. It contains $n$ wetted sites,
$b$
boundary sites which are already known to be non-wetted, and $g$ boundary sites
at which the cluster can still grow since their status has not yet been decided
(``growth sites"). This cluster will contribute to the percolation ensemble
only
if growth actually stops at all growth sites, i.e. with weight $(1-p)^g$. Since
the relative weights of the percolation and animal ensembles differ by a factor
$(1-p)^{(b+g)}$ (since now $b+g$ is the total number of boundary sites), this
cluster has weight $w(\cC)\propto (1-p)^{-b}$ in the animal ensemble. If we
would
use only this weight as a guide for cloning, we would clone if $w(\cC)$ is
larger
than some $W_+$ which is independent of $b$ and $g$, and which depends on $n$
in such a way that the sample size becomes independent of $n$.
But clusters with many growth sites will of course have a bigger chance to keep
growing and will contribute more to the precious statistics of very large
clusters.
It is not a priori clear what is the optimal choice for $W_+$ in view of this,
but
numerically we obtained best results for $W_+ \propto (1-p)^g$.

In this way we were able to obtain good statistics for animals of several
thousand
sites, independent of the dimension of the lattice. A typical 2-$d$ animal with
8000 sites is shown in Fig.~8. We were also able to simulate animal collapse
(when
each nearest neighbor pair contributes $-\epsilon$ to the energy), and animals
near an adsorbing surface. Details will be published in Ref.\cite{animals}.

\section{Implementation Details}

In this section, the notation will be appropriate for the lamb-and-lions
problem
of Sec.2, but all statements hold mutatis mutandis also for the other problems.

\subsection{Depth First Versus Breadth First}

As described in trick {\bf 1} of Sec.2, original enrichment was implemented ``breadth
first". There one keeps many replicas of the process simultaneous in the
computer, and advances them simultaneous. This is also the traditional way
of implementing evolutionary algorithms \cite{rechenberg,holland}. There it is
required for two reasons: Because of cross-over moves where two configurations
(``replicas", ``instances",
``individua", ...) are combined to give a new configuration, and because of
tournament selection where the less fit of a randomly selected pair is killed
and replaced by the more fit. In the present case there are no cross-overs, and
tournament selection is replaced by comparing the ``fitness" $w$ against
thresholds $W_\pm$ which will be determined by some average fitness.

This allows us to use a ``depth first" implementation where only a single
replica is kept in computer memory at any given time, and only when this is
pruned or has reached its final time $t_{\rm max}$, a new replica is started.
The names ``breadth first" and ``depth first" come from searching rooted trees
where the first searches the tree in full breadth before increasing the
depth, while in the second one first follows a branch in full depth and only
then considers alternative branches \cite{tarjan,sedgewick}.

The main advantages of depth first algorithms are reduced storage requirements
(which can be important even in present days of parallel machines where breadth
first algorithms can be implemented by putting each configuration on its own
node) and elegance of programming. While the `natural' coding paradigmes for
breadth first algorithms are iterations and first-in/first-out queues,
for depth first approaches they are recursions and stacks. In order to
implement the lamb-and-lions problem we need just a recursive function
STEP(t,w,x,x$_1$, ... x$_N$) whose arguments have the obvious meaning. When
called, it increases t $\to$ t$+1$, selects new positions from the neighbours
of the previous ones, updates w accordingly, and calls itself either twice
(cloning), once (normal evolution) or not at all (pruning). A pseudocode for
this is given in \cite{perm}.

There is a folklore saying that recursions are inefficient in terms of CPU
time \cite{thompson}, and large recursion depths should be avoided. It can be
avoided since
each recursion can also be re-coded as an iteration (FORTRAN 77, e.g., has no
recursion and is yet a universal language). But the speed-up is negligible on
modern compilers (less than 10\%, typically), depths of $10^4 - 10^5$ make no
problems, and readability of the code is much worse if recursion is not used.
We see only one reason for avoiding recursion, and that is its less efficient
use of main memory. In very large problems where stack size limitations can be
crucial, recoding in terms of iterations might be needed.

\subsection{Choosing $W_\pm$}
\label{wpm.sec}

In general, thresholds for pruning/cloning should not be too far apart since
otherwise the weights fluctuate too much and most of the total weight is
carried by few configurations only. We had best experiences with
$3 < W_+/W_- < 10$ in most applications, but other authors \cite{prellberg}
report good results also for $W_+/W_- \approx 100$. Obviously, the precise
value
is not very important.

More important is $W_+$ itself. As a rule of thumb, it should be chosen such
that the total number of configurations $C(t)$ created at time $t$ is
independent of $t$.
If $C(t)$ decreases with $t$, most of the CPU time is spent on small $t$
and the statistics at large $t$ depend on only few realizations. Inversely,
if $C(t)$ increases with $t$, all configurations at large $t$
are descendants of only few ancestors and are thus strongly correlated.

There is a very simple way to guarantee the approximate (up to a factor $\sim
1$)
constancy of $C(t)$ \cite{perm}. Let us denote by $Z(t)$ the path integral (or
partition sum), and $\hat{Z}(t)$ its estimate from the current simulation,
\be
   \hat{Z}(t) = M^{-1} \sum_j w_j(t)\;.
\ee
Here $M$ is the number of starting configurations which have already been
treated,
and $w_j(t)$ is the weight of the $j$-th configuration at time $t$. $C(t)$ will
be roughly independent of $t$ if
\be
   W_\pm = c_\pm \hat{Z}(t)        \label{wpl}
\ee
with $c_\pm$ being constants of order unity (typically, $c_+ \approx
1/c_-\approx [W_+/W_-]^{1/2}$).
We thus start the simulation with some guess for $W_\pm$ (the precise values
are
largely irrelevant, any large values would also do) and replace them by
Eq.~(\ref{wpl})
as soon as there was already a configuration at $t$, i.e. as soon as we have
$\hat{Z}(t)>0$.

More sophisticated ways \cite{fold,stiff} to choose $W_\pm$ are needed only for
very
hard problems with excessive pruning and cloning. In this case, the above
method would occasionally give excessively large ``tours" (a tour is
the set of all configurations which descend from the same ancestor, i.e. which
are obtained by cloning from the same starting configuration). To cut them
short, one should make $W_\pm$ larger than given by Eq.~(\ref{wpl}) if a tour
is already very large. We should however warn the reader that in such hard
cases
the estimates of partition sums are no longer reliable, and results should
be taken with some suspicion.

\subsection{Choosing the Bias}

As a general rule, the bias should be such that the bias correction factor
cancels exactly the Boltzmann weight (if there is one) and minimizes the
number of pruning/cloning events. A bad choice of the bias is immediately
seen in an increase of these events, and in a decrease of the number of tours
which reach large values of $t$. In $t$, a simulation corresponds essentially
to a random walk with reflecting boundary at $t=0$. While normal evolution
steps correspond to forward steps in $t$, pruning events correspond to
backward jumps to the last previous cloning time. A proper choice of $W_\pm$
eliminates any drift from this random walk, while a good bias maximized the
effective diffusion constant. If $W_+$ is chosen according to Eq.~(\ref{wpl}),
the CPU time needed to create an independent configuration at large $t$
increases essentially $\sim t^2$, the prefactor can be substantially
decreased by choosing a good bias.

Unfortunately, there is no universal recipe for such a good bias. There
is a general prescription in the case of diffusion quantum Monte Carlo
(see next section) and for related Markov processes, but even this is
usually not easy to implement. In other cases such as polymers with
self avoidance one has only heuristics. Sometimes even the algorithm
without bias is already very efficient, such as for multiple spanning
percolation clusters. In other cases, as in the lamb-and-lions problem,
the qualitative properties of the bias are obvious, but for its quantitative
implementation we had used trial and error.

One possible way to determine a good bias (or ``guiding", as it is sometimes
called) is to look $k$ steps ahead. For a polymer, e.g., one might try
all extensions of the chain by $k$ monomers, and decides on the success of
these extensions which single step to take next. This {\it scanning method}
\cite{meiro} is efficient in guiding the growth, but also very time consuming:
The effort increases exponentially with $k$. For polymers at low temperatures,
where Boltzmann factors can become very large, this may be efficient
nevertheless. But for a-thermal SAWs and for lattice polymers in the open coil
phase, another method is much more efficient.

\subsubsection{Markovian Anticipation}

In this alternative strategy \cite{cylinder} to guide polymer grow, one
essentially does
not look forward $k$ steps, but backward. Thus we remember during the growth
the last $k$ steps. On a lattice with coordination number $\cn$, this means
we label the present configuration by an integer $i = 1,\ldots \cn^k$.
Assume now that the next
step is in direction $j,\; j=1,\ldots\cn$. During the initial steps of the
simulation (or during an auxiliary run) we build up a histogram $H(i,j)$ of
size $\cn^{(k+1)}$. There we add up the weights with which all configurations
with history $(i,j)$ between the steps $n-k$ and $k$ contribute to the
partition sum
of chains with length $n+m$, with $m \gg 1$ (we typically use $m \approx 100 -
200$).
The ratio
\be
   \hat{p}(j|i) = { H(i,j) \over \sum_{j'} H(i,j')}            \label{ma}
\ee
is then an estimate of how efficient the extension $j$ was in the long run.
After some obvious modifications taking into account that there is no history
yet for the first $k$ steps, and that no anticipation is useful for the last
few steps, we use $\hat{p}(j|i)$ (which is properly normalized already!) as the
probability with which we make step $j$, given the history $i$.

Note that this can also be used, e.g., for stretched polymers where the
$\hat{p}(j|i)$ are not isotropic, and where one can anticipate that the next
monomer should be added preferentially in the direction of stretching.

\subsection{Error Estimates and Reliability Tests}

Errors can in principle be estimated {\it a priori} and {\it a posteriori}. In
the
former case one knows them even before making the simulations. For instances,
if one draws $n$ realizations of a random number with variance $\sigma$, the
average has variance $\sigma/N$. A posteriori errors, in contrast, are obtained
from fluctuations between the different realizations.

A priori errors for go-with-the-winners simulations are possible \cite{aldous}
but difficult because the generated sample is correlated. Indeed, making such
estimates was the main objective of \cite{aldous}, but the compromizes as
regards
efficiency are such that the results obtained there seem not very practical.

A posteriori errors can be made easily by dividing a long run into several
bunches, computing averages over each bunch, and studying the fluctuations
between
them. This is essentially also the strategy in standard Metropolis simulations,
but here the situation is even simpler. Since each `tour' (see
Sec.\ref{wpm.sec})
is independent from any other, the break-up into bunches just has to be between
tours. No problem due to correlations of uncertain range as in Metropolis
simulations occurs here.

Nevertheless, the problems of critical slowing down and of being trapped in
local free energy minima which plague Metropolis simulations are not absent
in go-with-the-winners simulations. They just appear in new guises. Namely,
single tours can become extremely large. If that happens, nearly the entire
weight
accumulated during a long simulation can be carried by a single tour or,
what is even worse, the tours of {\it really} large weight have not been
found at all. The latter is the analogon to not yet having reached
equilibrium in Metropolis simulations.

Although this is first of all a problem of error bars, it can easily, if
one is not very careful, turn into a source of systematic errors. This is
because one is primarily not interested in the partition sum (which is
always sampled without a bias), but in its logarithm or in derivatives thereof.
Consider e.g. a situation where we want to make several independent runs
since we want to make sure that we have made everything right. From
each simulation we estimate a free energy, and then we take their average
value as our final estimate. If the problem is really hard, the fluctuations
of the partition sums will be non-Gaussian, with very many small downward
fluctuations compensated by few large positive fluctuations. By taking the
logarithm, the latter are cut down, and a negative bias results.

There is no fool-proof remedy against this danger. But there is an
easy and straightforward way to check that at least that part of phase
space which has been visited at all has been sampled sufficiently during
a single run. For this, we make a histogram of tour weights on a logarithmic
scale, $P(\log(w))$, and compare it with the weighted histogram $wP(\log(w))$.
If the latter has its maximum for values of $\log(w)$ where the former is
already large (i.e. where the sampling is already sufficient), we are
presumably on the safe side. However, if $wP(\log(w))$ has its maximum
at or near the upper end of the  sampled range, we should be skeptical.

As an example, we show results for a self avoiding 2-d walk in a random
medium \cite{localize}. This medium is an infinite square lattice with (frozen)
random energies $E_i$ on each site $i$. In particular, $E_i$ is either $-1$
(with probability $p$) or 0 (with probability $1-p$). The polymer is free to
float in the entire lattice. Previous simulations \cite{baumg} had
suggested that for any finite $p$ there is a phase transition at $T=T_c(p)$,
maybe because the polymer becomes localized in an ``optimal" part of the
lattice for $T<T_c$.

\begin{figure}
\begin{center}
\parbox{.52\textwidth}{\includegraphics[width=.40\textwidth,angle=270]{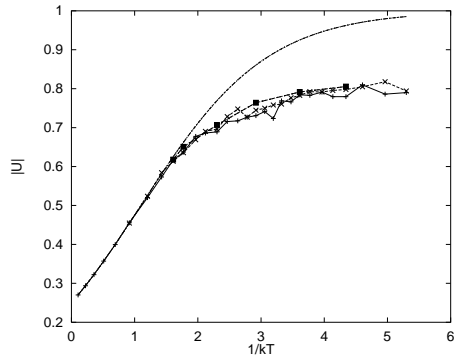}}
\parbox{.42\textwidth}{\caption[]{Absolute value of $U$ for $p=1/4$ and
$N=200$. The
continuous line is the exact theoretical result.
Pluses $(+)$ have low statistics (ca. 10 to 20 min CPU time
per point); crosses $(\times)$ have medium statistics (about a factor 10 more);
squares have about a factor 20 more statistics than crosses.}}
\end{center}
\label{fig15}
\end{figure}

\begin{figure}
\begin{center}
\parbox{.49\textwidth}{\includegraphics[width=.40\textwidth,angle=270]{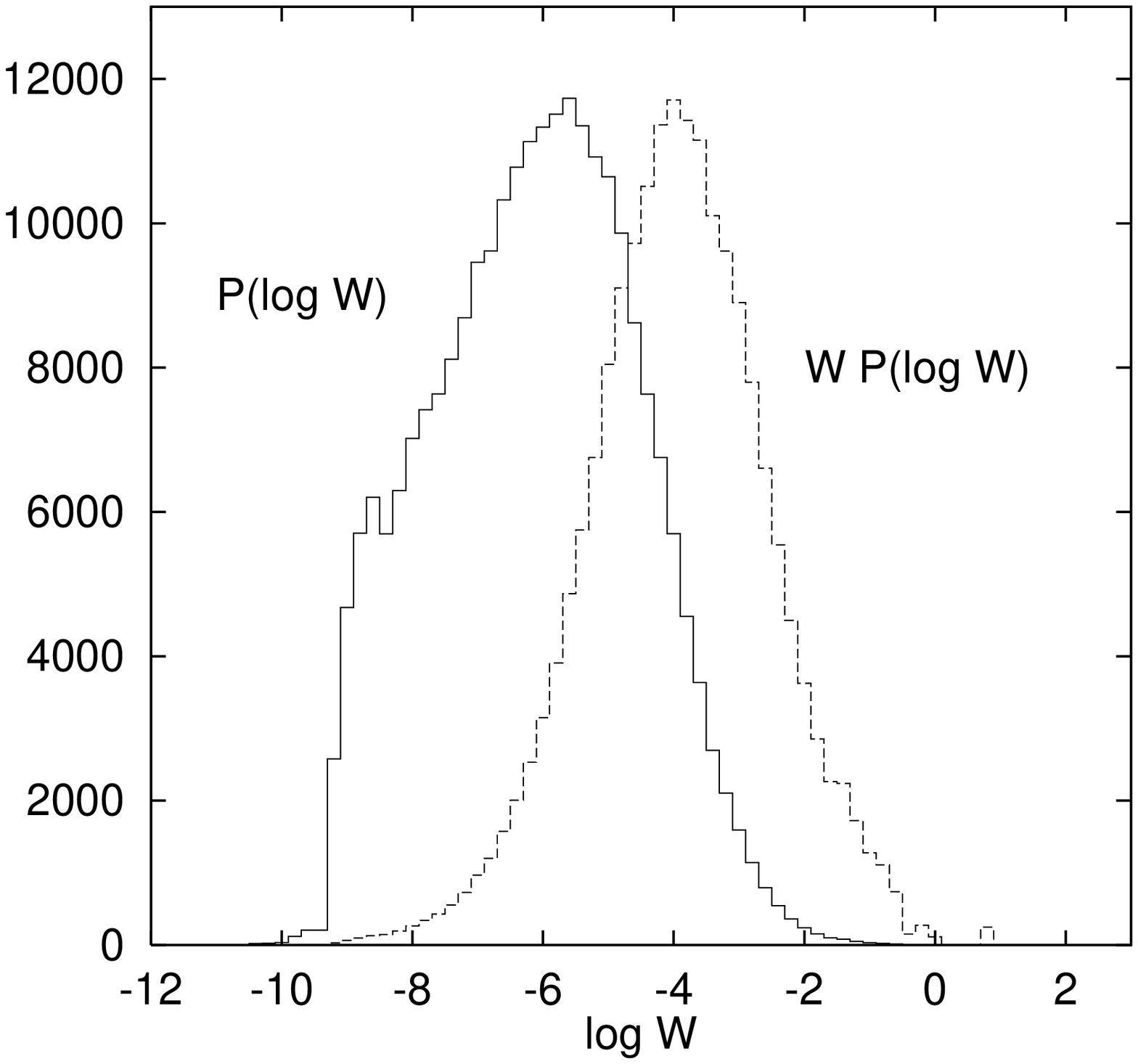}}
\parbox{.49\textwidth}{\includegraphics[width=.40\textwidth,angle=270]{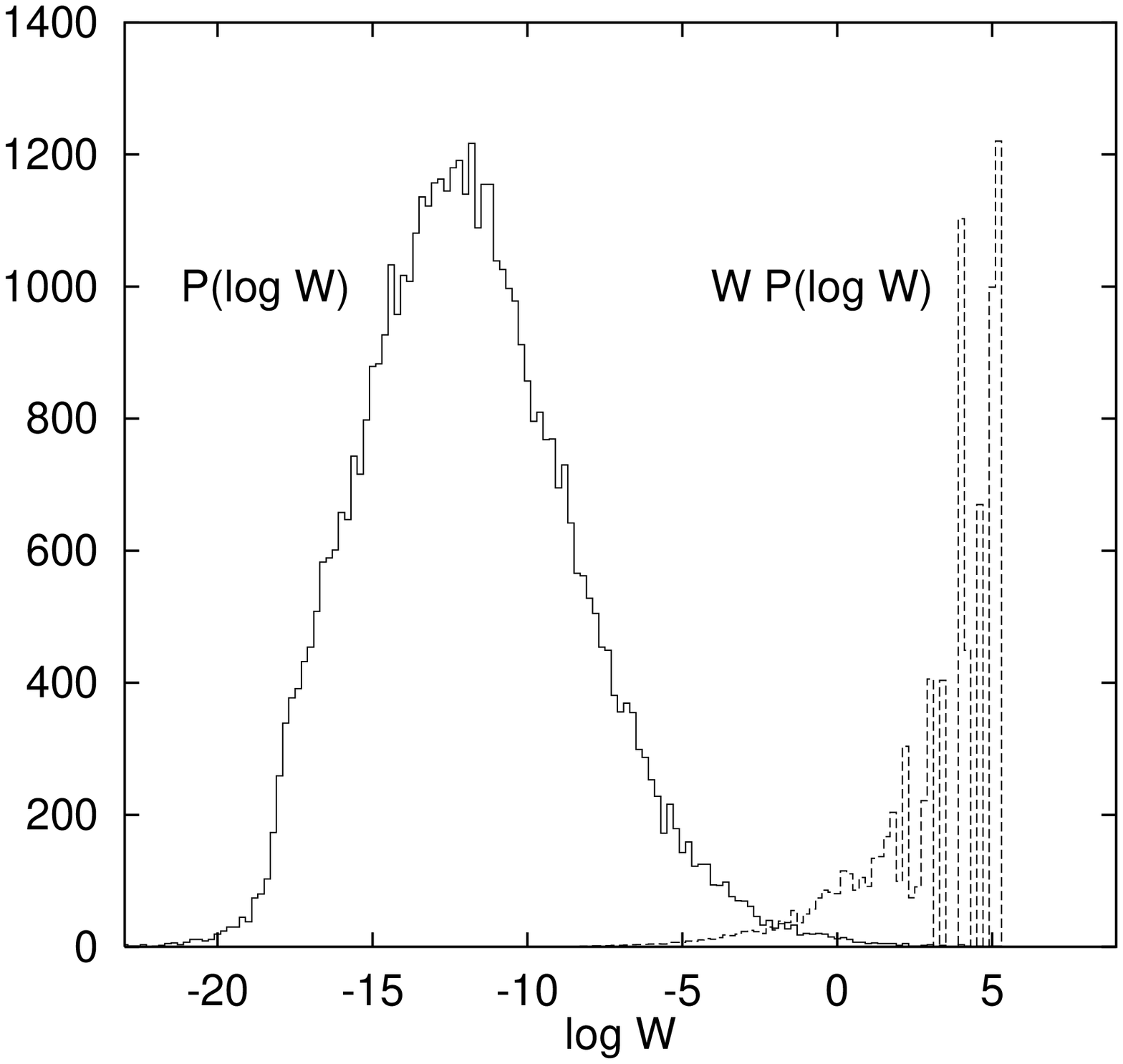}}
\caption[]{Full lines are histograms of logarithms of tour weights for
$1/kT= 0.92$ (left panel) resp. 2.30 (right), normalized as
tours per bin.  Broken lines show the corresponding weighted distributions,
normalized so as to have the same maximal heights. Weights $W$ are only
fixed up to a $\beta$-dependent multiplicative constant.}
\end{center}
\label{fig1617}
\end{figure}

But one can show rigorously that no such transition can exist \cite{localize}.
In order to understand the source of the problem, we made simulation with
PERM and monitored the distribution of tour weights. Results are shown in
Figs. 9 and 10. In Fig.9 we show the average values of the
energy\footnote{Here,
$\cC$ denotes the set of sites occupied by the walk, and averaging is done
over all walks and all disorder realizations},
$|U| = - \langle \sum_{i\in \cC} E_i\rangle$ for $p=1/4$ and chain length
$N=200$. Three curves obtained from simulations are shown together with
a curve obtained analytically. The fact the three curves, obtained with
vastly different statistics, agree with each other but deviate from the
theoretical curve at the same value of $T$ indicates the supposed
phase transition. But histograms obtained in the regions where theory
and simulations (dis-)agree (see Fig.10) show clearly that the simulations
for $1/kT>2$ are not reliable (right panel) while those for $1/kT<1.5$ are.

\section{Diffusion Quantum Monte Carlo}

For completeness we sketch here shortly the main idea of diffusion type
quantum MC (QMC) simulations \cite{kalos,anderson,vonderlinden},
to show how they are related to the previous
calculations and to understand why a perfect bias is in principle possible
in QMC but not in classical applications.

We start from the simplest version of a time dependent Schr\"odinger equation,
\be
   i \partial\psi(x,t)/\partial t = -(2m)^{-1} \nabla^2\psi(x,t) + V(x)
\psi(x,t),
\ee
and we are interested in finding its ground state energy, i.e.
we want to solve the time-independent Schr\"odinger equation
\be
    E_{\rm min}\psi(x) = -(2m)^{-1} \nabla^2\psi(x) + V(x) \psi(x),  \label{s2}
\ee
for the smallest eigengenvalue $E_{\rm min}$.
Replacing $t$ by an imaginary ``time" and $(2m)^{-1}$ by a diffusion
coefficient $D$
we end up at a diffusion equation
\be
    \partial\psi(x,t)/\partial t = D \nabla^2\psi(x,t) - V(x) \psi(x,t)
  \label{diff}
\ee
with an external source/sink $V(x)$ and $\psi$ viewed as a classical density.
The ground state energy of the original problem is now transformed into the
slowest relaxation rate.
If we want to simulate this by diffusing particles, we can take the last
term into account by either killing particles (if $V >0$) and cloning them
(if $V<0$), or by giving them a weight $\exp[\int dt V(x(t))]$. Neither is
very efficient. For efficiency, we should rather replace the random walk
by a biased (``guided") motion for which neither weighting nor killing/cloning
is needed.

For this purpose we choose a ``guiding function" $g(x)$ and write
\be
    \psi(x,t) = \rho(x,t)/g(x)\;.
\ee
Eq.(\ref{diff}) leads then to the following equation for $\rho$:
\be
    \partial\rho/\partial t = D \nabla^2\rho - [V(x)-D{\nabla^2g(x)\over g(x)}]
\rho
               - \nabla\left([2D{\nabla g(x)\over g(x)}] \; \rho\right)
  \label{phi}
\ee
This is now a diffusion equation with drift (last term) and with a modified
source/sink term.  If the latter is constant, i.e.
\be
   D\nabla^2 g(x) - V(x)\: g(x) = {\rm const}\; g(x),      \label{g}
\ee
then no killing/pruning is needed and the weight increase/decrease is uniform
and thus trivial. But Eq.~(\ref{g}) is just the time-independent Schr\"odinger
equation, Eq.~(\ref{s2}), we wanted solve. It seems that we have gained
nothing. For an
optimal implementation, we have to know already the solution we want to get.

Things are of course not so bad since we can proceed iteratively: start with a
rough guess for $g(x)$, obtain with it an estimate for $\psi(x)$, use it as the
next guess for $g(x)$, etc.

A crucial observation now is the following: the density $\rho(x)$ of the
guided diffusers is, if $g(x)$ satisfies Eq.~(\ref{g}), just equal to the
quantum mechanical density\footnote{Note that $\psi(x)$ is real here since
we are interested in the ground state. However, a similar derivation is
possible
when using a time-dependent guiding function $g(x,t)$. Then Eq.~(\ref{g})
becomes the adjoint time-dependent Schr\"odinger equation.},
 $|\psi(x)|^2$.
Thus random sampling of Eq.~(\ref{phi})
corresponds precisely to random sampling of the quantum-mechanical density.
We thus really have solved the problem of importance sampling.

Note that if we had {\it started} instead with Eq.~(\ref{diff})
as a {\it classical} problem, and
were interested in the {\it classical} density, we would not make perfect
importance sampling: the particles would be sampled in the simulation
not with the density they should have. Although using Eq.~(\ref{g}) for the
guiding function would still be formally correct, it would not lead to minimal
statistical fluctuations.

\section{Conclusion}

We have seen that stochastic simulations {\it not} following the traditional
Metropolis scheme can be very efficient. We have illustrated this with a
wide range of problems. Conspicuously, the Ising model was not among them.
The reason is simply that no go-with-the-winners algorithm for the Ising
model has been proposed which is more efficient than, say, the Swendsen-Wang
\cite{swendsen} algorithm. But there is no reason why such an algorithm should
not exist. In principle, the go-with-the-winners strategy has at least as wide
a range of applications as the Metropolis Metropolis scheme. Its only
requirement
is that instances (configurations, histories, ...) are built up in small steps,
and that the growth of their weights during the early steps of this build-up is
not
too misleading.

The method is not new. It has its roots in algorithms which are regularly used
since several decades. Some of them, like genetic algorithms, are familiar to
most scientists, but it is in general not well appreciated that they can be
made into a general purpose tool. And it seems even less appreciated how
closely methods developed for quantum MC simulations, polymer simulations, and
optimization methods are related. I firmly believe that this close relationship
can be made use of in many more applications to come.

\vspace{.4cm}
Acknowledgements: We are indebted to Drs. Rodrigo Quian Quiroga and G\"unther
Radons for discussions and for carefully reading the manuscript. WN is supported 
by DFG, SFB 237.

%
\end{document}